\documentstyle[subeqn]{article}

\input amssym.def
\input amssym.tex

\def\ja{j}

\def\Q1{Q}
\def\qa{{\cal Q}}

\def\QS1{{\sf q}}

\def\R1{R}
\def\ra{{\cal R}}

\def\RS1{{\sf r}}

 \def\eq{\hspace{-2.5mm}&=&\hspace{-2.5mm}}
 \def\espace{\hspace{-2.5mm}&&\hspace{-2.5mm}}

\def\tz{z}

\def\beq{\begin{equation}} \def\eeq{\end{equation}}
\def\bseq{\begin{subequations}} \def\eseq{\end{subequations}}
\def\bea{\begin{eqnarray}} \def\eea{\end{eqnarray}}
\def\bsea{\begin{subeqnarray}} \def\esea{\end{subeqnarray}}

\let\ti=\tilde  

\let\nn=\nonumber
\def\beann{\begin{eqnarray*}} \def\eeann{\end{eqnarray*}}

\let\a=\alpha \let\be=\beta  \let\de=\delta

 \let\Ds=\displaystyle 

\newcommand{\eins}{\mbox{1 \hspace{-9.2pt} I}}

\def\0{\over } \def\1{\vec }     \def\2{{1\over2}} \def\4{{1\over4}}
\def\5{\bar }  \def\6{\partial } \def\7#1{{#1}\llap{/}}

\def\<{\langle } \def\>{\rangle }

\def\i{{\rm i}}

\def\d{{\rm d}}

\def\e{{\rm e}}

 \def\sech{\mbox{\,sech}}

\title{
Integrable Discretization of the Coupled Nonlinear Schr\"{o}dinger Equations
}
\author{Takayuki Tsuchida\thanks{ \hspace*{0.5mm} Supported by JSPS Research 
Fellowships for Young Scientists.} \\ 
Department of Physics, Graduate School of Science, \\
University of Tokyo, 
Hongo 7--3--1, Bunkyo-ku, \\
Tokyo 113--0033, Japan \\ 
tsuchida@monet.phys.s.u-tokyo.ac.jp \\[2ex]}

\begin{document}

\maketitle
\begin{abstract}
A discrete version of the inverse scattering 
method proposed by Ablowitz and Ladik is generalized to 
study an integrable full-discretization (discrete time and 
discrete space) of the coupled nonlinear 
Schr\"{o}dinger equations. The generalization enables 
one to solve the initial-value problem. Soliton solutions 
and conserved quantities of the full-discrete system are 
constructed. 
\end{abstract}

\section{Introduction}

Among various soliton equations, a system of coupled nonlinear 
Schr\"{o}dinger (CNLS) equations,
\beq
\begin{array}{l}
\Ds \i \frac{\partial q_j}{\partial t}
 + \frac{\partial^2 q_j}{\partial x^2}
- 2\sum_{k=1_{\vphantom \mu}}^m q_k r_k \cdot q_j
=0,\vphantom{\sum_{\stackrel{B}{<}}}
\\
\Ds \i \frac{\partial r_j}{\partial t}
 - \frac{\partial^2 r_j}{\partial x^2}
+ 2\sum_{k=1}^m r_k q_k \cdot r_j =0,
\end{array}
\hspace{10mm}
j=1,2, \ldots, m,
\label{CNLS}
\eeq
is physically significant since it describes multi-mode 
wave propagation in nonlinear optics. The CNLS equations can be 
solved via the inverse scattering method (ISM) under the reduction 
$r_j = - q_j^{\ast}$ and the decaying boundary conditions 
$q_j \to 0$ as $x \to \pm \infty$. It is, however, not always 
easy to trace time evolution of initial values for soliton equations 
solvable via the ISM\@. Thus, it is an important problem to find 
good schemes for numerical computation of soliton equations. One 
possible approach is to find discretization of soliton equations 
with preserving the complete integrability (integrable 
discretization for short). The discretized soliton equations are not only 
useful for numerical computation but also interesting as models of
nonlinear population dynamics, nonlinear electric
circuits and nonlinear lattice vibrations.

An integrable discretization of the nonlinear Schr\"{o}dinger (NLS) equation 
((\ref{CNLS}) with $m=1$) was proposed by Ablowitz and Ladik \cite{AL1,AL2}. The author, Ujino and Wadati generalized the formulation 
of Ablowitz and Ladik and found an integrable semi-discretization (space 
discretization) of the coupled modified KdV equations \cite{Tsuchida2} 
and the CNLS equations \cite{Tsuchida3}. In this article, as a 
continuation of \cite{Tsuchida3}, we consider an integrable
discretization of both space and time (full-discretization for short) 
for the CNLS equations (\ref{CNLS}). In accordance with the page
limit, we concisely explain an outline of the ISM for the model. 
The reader can refer to \cite{Tsuchida2,Tsuchida3} for more 
details because we can perform the ISM in the same manner as in these 
papers, apart from the difference in time dependences of scattering data.

\section{Lax pair}

We introduce a set of auxiliary linear equations,
\beq
\Psi_{n+1} = L_n \Psi_n,
\hspace{5mm} \ti{\Psi}_n = V_n \Psi_n,
\label{aux}
\eeq
where the tilde $(\; \ti{} \; )$ denotes the time shift in discrete
time $l \in {\Bbb Z}$ ($l \to l+1$). 
The compatibility condition of (\ref{aux}) is given by
\beq
\ti{L}_n V_n = V_{n+1} L_n .
\label{Lax_full}
\eeq
$L_n$ and $V_n$, and (\ref{Lax_full}) are respectively 
the Lax pair and the zero-curvature condition (or Lax equation) 
in the full-discrete case. 
To obtain a full-discrete version of the CNLS equations, 
we choose the form of the Lax pair as
%
%
\bea
L_n  \eq
z \left[
\begin{array}{cc}
 \e^{\i \a H_1}  &  \\
  &  O \\
\end{array}
\right]
+ 
\left[
\begin{array}{cc}
O  & \e^{\i \a  H_1} \qa_n \\
\e^{\i \a H_2} \ra_n  & O \\
\end{array}
\right]
+ \frac{1}{z}
\left[
\begin{array}{cc}
 O &   \\
   & \e^{\i \a H_2} \\
\end{array}
\right],
\label{L_form}
\eea
\bea
V_n \eq 
\left[
\begin{array}{cc}
 \e^{\i \be H_1}  &  \\
   & \e^{\i \be H_2} \\
\end{array}
\right]
%
+ \frac{ \sin \be}{\sin (2\a +\be)} \ja_n 
\Biggl\{
z^2 
\left[
\begin{array}{cc}
O &   \\
  & -\e^{\i \be H_2} \\
\end{array}
\right]
\nn \\
\espace 
%
\mbox{}
+ z
\left[
\begin{array}{cc}
 O & \ti{\qa}_n \e^{\i \be H_2} \\
 \e^{\i (\a + \be)H_2} \ra_{n-1}\e^{-\i \a H_1} & O \\
\end{array}
\right]
\nn \\
\espace 
\mbox{} +
\left[
\begin{array}{cc}
- \ti{\qa}_n \e^{\i (\a+\be)H_2} \ra_{n-1}\e^{-\i \a H_1} &   \\
 & - \ti{\ra}_n \e^{\i (\a + \be) H_1} \qa_{n-1}
 \e^{-\i \a H_2} \\
\end{array}
\right]
\nn \\
\espace 
%
%
\mbox{} + \frac{1}{z}
\left[
\begin{array}{cc}
 O  & \e^{\i(\a+\be)H_1} \qa_{n-1} \e^{-\i \a H_2} \\
 \ti{\ra}_n \e^{\i \be H_1} & O \\
\end{array}
\right]
+ \frac{1}{z^2}
\left[
\begin{array}{cc}
 -\e^{\i \be H_1} &  \\
  &  O \\
\end{array}
\right]  \Biggr\}.
\label{V_form}
\eea
Here $z$ is the spectral parameter which is time-independent. 
$\qa_n$ and $\ra_n$ are $p \times p$ matrices. 
$H_1$ and $H_2$ are $p \times p$ constant matrices. $\a$ and 
$\be$ are constants. $j_n$ is a scalar variable. 
We assume the following relations,
\beq
[H_1, \qa_n \ra_n] = [H_2, \ra_n \qa_n] =O,
\label{H_qr}
\eeq
\beq
H_1 \qa_n - \qa_n H_2 = -2 F_1 \qa_n F_2,
\hspace{5mm}
H_2 \ra_n - \ra_n H_1 = 2 F_2 \ra_n F_1,
\label{HqrF}
\eeq
where $[ \, \cdot \, , \, \cdot \, ]$ denotes the commutator and 
$F_1$, $F_2$ are $p \times p$ constant matrices which satisfy
\beq
(F_1)^2 = (F_2)^2 = I,
\hspace{5mm}
[F_1, H_1] = [F_2, H_2] = O.
\label{F_H}
\eeq
Here $I$ is the identity matrix. We notice that the following relations hold:
\bseq
\bea
\e^{-\i y H_1} \qa_n \e^{\i y H_2} = 
\cos (2y) \qa_n + \i \sin (2y) F_1 \qa_n F_2 ,
\\
\e^{\i y H_2} \ra_n \e^{- \i y H_1} = 
\cos (2y) \ra_n + \i \sin (2y) F_2 \ra_n F_1 .
\eea
\label{ex}
\eseq
We write the difference interval of time by $\de t$. 
Putting (\ref{L_form}) and (\ref{V_form}) into (\ref{Lax_full}), 
we obtain
\bseq
\bea
&& \hspace{-10mm}
\i\frac{1}{\de t}(\ti{\qa}_n-\qa_n) 
-\i \frac{\tan \be \cot (2\a + \be)}{\de t }
\{j_{n+1} \ti{\qa}_{n+1}(I- \ra_n \qa_n)
-j_{n+1} (I-\qa_n \ra_n)\qa_{n-1}\}
\nn \\
&& 
\hspace{-10mm}
\mbox{} + \frac{\tan \be}{\de t}
\{ \ja_{n+1} F_1 \ti{\qa}_{n+1} F_2 (I - \ra_n \qa_n)
+\ja_{n+1} (I-\qa_n \ra_n) F_1 \qa_{n-1}F_2 
\nn \\
\Ds
&& 
\hspace{-10mm} 
\mbox{} - F_1 (\ti{\qa}_n + \qa_n) F_2 \}  =O,
\vphantom{\sum_{k}}
\label{mat1}
\\
&& \hspace{-10mm}
\i\frac{1}{\de t}(\ti{\ra}_n-\ra_n) 
-\i \frac{\tan \be \cot (2\a + \be)}{\de t}
\{ j_{n+1} \ti{\ra}_{n+1} (I-\qa_n \ra_n)
-j_{n+1} (I-\ra_n \qa_n ) \ra_{n-1} \}
\nn \\
&& 
\hspace{-10mm}
\mbox{} - \frac{\tan \be}{\de t}
\{ \ja_{n+1} F_2 \ti{\ra}_{n+1} F_1 (I - \qa_n \ra_n)
+\ja_{n+1} (I-\ra_n \qa_n) F_2 \ra_{n-1}F_1 
\nn  \\
&& 
\hspace{-10mm} 
\Ds
\mbox{} - F_2 (\ti{\ra}_n + \ra_n) F_1 \} =O,
\vphantom{\sum}
\label{mat2} 
\\
&&  \hspace{20mm} 
\ja_{n+1} (I-\qa_n \ra_n) = \ja_n (I-\ti{\qa}_n \ti{\ra}_n),
\vphantom{\sum}
\label{ja1}
\\
&&  \hspace{20mm} 
\ja_{n+1} (I-\ra_n \qa_n) = \ja_n (I-\ti{\ra}_n \ti{\qa}_n),
\label{ja2}
\eea
\eseq
with the help of (\ref{ex}). 
If we set
\[
2\a+\be = \frac{\pi}{2}, \hspace{5mm} 
\frac{\tan \be}{\de t} = 1,
\]
or equivalently,
\[
\a = \frac{\pi}{4}-\frac{1}{2} \tan^{-1} \de t , 
\hspace{5mm} \be = \tan^{-1} \de t ,
\]
(\ref{mat1}) and (\ref{mat2}) are simplified as
\bseq
\bea
&& \hspace{-7mm}
\i\frac{1}{\de t}(\ti{\qa}_n-\qa_n) 
+ \ja_{n+1} F_1 \ti{\qa}_{n+1} F_2 (I - \ra_n \qa_n)
+ \ja_{n+1} (I-\qa_n \ra_n) F_1 \qa_{n-1}F_2 
\nn \\
&& \hspace{-7mm}
\mbox{} - F_1 (\ti{\qa}_n + \qa_n) F_2  =O,
\vphantom{\sum_{k}}
\label{mat3}
\\
&& \hspace{-7mm}
\i\frac{1}{\de t}(\ti{\ra}_n-\ra_n) 
-  \ja_{n+1} F_2 \ti{\ra}_{n+1} F_1 (I - \qa_n \ra_n)
-  \ja_{n+1} (I-\ra_n \qa_n) F_2 \ra_{n-1}F_1 
\nn \\
&& \hspace{-7mm}
\mbox{} + F_2 (\ti{\ra}_n + \ra_n) F_1 =O.
\label{mat4}
\eea
\eseq
To consider a reduction of these matrix equations to full-discrete 
CNLS equations, we define $F_1$, $F_2$, $H_1$, $H_2$, $\qa_n$ and 
$\ra_n$ recursively as follows:
 \[
 F^{(1)}_1 = 1, \hspace{3mm} F^{(1)}_2 = 1, \hspace{3mm}
 H_1^{(1)} = -1, \hspace{3mm} H_2^{(1)} = 1, \hspace{3mm}
 \qa^{(1)}_n = q^{(1)}_n, \hspace{3mm} \ra^{(1)}_n = r^{(1)}_n,
 \]
 \[
 F^{(m+1)}_1
 = 
 \left[
 \begin{array}{cc}
  F^{(m)}_1 &   \\
    & - F^{(m)}_2 \\
 \end{array}
 \right],
\hspace{5mm}
 F^{(m+1)}_2
 = 
 \left[
 \begin{array}{cc}
  F^{(m)}_2 &   \\
    &  F^{(m)}_1 \\
 \end{array}
 \right],
 \]
\[
 H^{(m+1)}_1
 = 
 \left[
 \begin{array}{cc}
  H^{(m)}_1 - I_{2^{m-1}} &  \\
    &    H^{(m)}_2 + I_{2^{m-1}} \\
 \end{array}
 \right],
\]
\[
 H^{(m+1)}_2
 = 
 \left[
 \begin{array}{cc}
  H^{(m)}_2 - I_{2^{m-1}} &  \\
    &  H^{(m)}_1 + I_{2^{m-1}} \\
 \end{array}
 \right],
\]
\[
 \qa^{(m+1)}_n
 = 
 \left[
 \begin{array}{cc}
  \qa^{(m)}_n &  q^{(m+1)}_n I_{2^{m-1}} \\
  r^{(m+1)}_n I_{2^{m-1}} &  -\ra^{(m)}_n \\
 \end{array}
 \right],
\hspace{5mm}
 \ra^{(m+1)}_n
 = 
 \left[
 \begin{array}{cc}
  \ra^{(m)}_n &  q^{(m+1)}_n I_{2^{m-1}} \\
  r^{(m+1)}_n I_{2^{m-1}} &  -\qa^{(m)}_n \\
 \end{array}
 \right].
 \]
Here $I_{2^{m-1}}$ is the $2^{m-1} \times 2^{m-1}$ identity 
matrix. We can easily show that (\ref{H_qr})--(\ref{F_H}) and the 
relation, 
 \[
 Q^{(m)}_n R^{(m)}_n = R^{(m)}_n Q^{(m)}_n 
        = \sum_{k=1}^{m} q^{(k)}_n r^{(k)}_n \cdot I_{2^{m-1}},
 \]
are satisfied. Then it is straightforward to prove that substitution 
of $F_1^{(m)}$, $F_2^{(m)}$, $\qa_n^{(m)}$ 
and $\ra_n^{(m)}$ into (\ref{ja1}), (\ref{ja2}), (\ref{mat3}) and 
(\ref{mat4}) yields the following full-discrete CNLS equations,
%
\bea
&& \hspace{-5mm}
\i \frac{1}{\de t} (\ti{q}_n^{(j)}-q_n^{(j)})
+ \Bigl(1- \sum_{k=1}^m q_n^{(k)} r_n^{(k)} \Bigr) \ja_{n+1} 
(\ti{q}_{n+1}^{(j)} + q_{n-1}^{(j)}) 
- (\ti{q}_n^{(j)}+q_n^{(j)}) = 0,
\vphantom{\sum_{\stackrel{B}{<}}}
\nn \\
&& \hspace{-5mm}
\i \frac{1}{\de t} (\ti{r}_n^{(j)}-r_n^{(j)})
- \Bigl(1- \sum_{k=1}^m r_n^{(k)} q_n^{(k)} \Bigr) \ja_{n+1} 
(\ti{r}_{n+1}^{(j)} + r_{n-1}^{(j)}) 
+ (\ti{r}_n^{(j)}+r_n^{(j)}) = 0,
\vphantom{\sum_{\stackrel{B}{<}}}
\nn \\
&& \hspace{-5mm}
\hspace{10mm}
\ja_{n+1} \Bigl( 1-\sum_{k=1}^m q_n^{(k)} r_n^{(k)} \Bigr) 
= \ja_n \Bigl( 1-\sum_{k=1}^m \ti{q}_n^{(k)} \ti{r}_n^{(k)} \Bigr) ,
\hspace{5mm} j=1,2,\ldots, m.
\label{fCNLS}
\eea
%
Assuming the reduction $r^{(j)}_n = - q^{(j) \, \ast}_n$, 
we obtain the self-focusing case of the full-discrete CNLS equations:
\bea
\espace 
\i \frac{1}{\de t} (\ti{q}_n^{(j)}-q_n^{(j)})
+ \Bigl(1 + \sum_{k=1}^m |q_n^{(k)}|^2 \Bigr) \ja_{n+1} 
(\ti{q}_{n+1}^{(j)} + q_{n-1}^{(j)}) 
- (\ti{q}_n^{(j)}+q_n^{(j)}) = 0,
\vphantom{\sum_{\stackrel{B}{<}}}
\nn \\
\espace 
\hspace{10mm}
\ja_{n+1} \Bigl( 1+\sum_{k=1}^m |q_n^{(k)}|^2 \Bigr) 
= 
\ja_n \Bigl( 1 + \sum_{k=1}^m |\ti{q}_n^{(k)}|^2 \Bigr) ,
\hspace{5mm} j=1,2, \ldots, m.
\label{fdcH}
\eea
In the one-component case ($m=1$), this system coincides with 
the full-discrete NLS equation obtained by Ablowitz and Ladik \cite{AL2}. 
If we take the continuum limit of time ($\de t \to 0$), we 
can equate $\ja_n$ to $1$. Thus, the system 
(\ref{fCNLS}) reduces to the semi-discrete CNLS 
equations \cite{Tsuchida3}, 
%
\bea
\begin{array}{l}
\Ds \i \frac{\partial q^{(j)}_n}{\partial t}
 + (q^{(j)}_{n+1}+q^{(j)}_{n-1}-2q^{(j)}_{n})
- \sum_{k=1}^m q^{(k)}_n r^{(k)}_n \cdot (q^{(j)}_{n+1}+q^{(j)}_{n-1}) 
=0,\vphantom{\sum_{\stackrel{B}{<}}}
\\
\Ds \i \frac{\partial r^{(j)}_n}{\partial t}
 - (r^{(j)}_{n+1}+r^{(j)}_{n-1}-2r^{(j)}_{n})
+ \sum_{k=1}^m r^{(k)}_n q^{(k)}_n \cdot (r^{(j)}_{n+1}+r^{(j)}_{n-1}) 
=0,
\vphantom{\sum_{\stackrel{B}{<}}}
\end{array}
\hspace{2mm}
j=1, 2, \ldots, m .
\nn
\eea
The corresponding linear problem is given by taking the 
continuum limit:
\[
\Psi_{n+1} = L_n \Psi_n,
\hspace{5mm} 
\frac{ \partial \Psi_{n}}{\partial t} = M_n \Psi_n,
\]
with
\bea
\hspace{-10mm}  L_n
\eq
  \tz \left[
  \begin{array}{cc}
    \e^{\i \frac{\pi}{4} H_1} &   \\
      &  O  \\
  \end{array}
  \right]
  +
  \left[
  \begin{array}{cc}
   O  &  \e^{\i \frac{\pi}{4} H_1} \qa_n \\
   \e^{\i \frac{\pi}{4} H_2} \ra_n  &  O \\
  \end{array}
  \right]
 +
  \frac{1}{\tz} \left[
  \begin{array}{cc}
    O  &   \\
      &  \e^{\i \frac{\pi}{4} H_2} \\
  \end{array}
  \right],
\nn
  \eea
  \bea
M_n \eq  \tz^2
  \left[
  \begin{array}{cc}
  O &   \\
     &  -I \\
  \end{array}
  \right]
  + \tz
  \left[
  \begin{array}{cc}
   O  &  \qa_n \\
   \e^{\i \frac{\pi}{4} H_2} \ra_{n-1} \e^{-\i \frac{\pi}{4} H_1} &  O  \\
  \end{array}
  \right]
\nn \\
\espace
\mbox{}
 +  \left[
  \begin{array}{cc}
 - \qa_n \e^{\i \frac{\pi}{4} H_2} \ra_{n-1} \e^{-\i
   \frac{\pi}{4}H_1}+\i H_1 \\
  & -\ra_n \e^{\i \frac{\pi}{4} H_1} \qa_{n-1} \e^{-\i
   \frac{\pi}{4}H_2}+\i H_2 \\
 \end{array}
 \right]
\nn \\
\espace
\mbox{}
 + \frac{1}{\tz}
  \left[
  \begin{array}{cc}
   O  & \e^{\i \frac{\pi}{4} H_1} \qa_{n-1} \e^{-\i \frac{\pi}{4} H_2} \\
   \ra_{n} &  O  \\
  \end{array}
  \right]
  + \frac{1}{\tz^2}
  \left[
  \begin{array}{cc}
   -I  &   \\
    & O \\
  \end{array}
  \right].
\nn
\eea
It should be noted that the above Lax pair for the semi-discrete 
CNLS equations is equivalent to the one given in \cite{Tsuchida3} 
via a gauge transformation.

\section{Inverse scattering method}
\label{}



As was shown in the previous section, the linear problem 
for the full-discrete CNLS equations 
(\ref{fCNLS}) is expressed in terms of a $2^m \times 2^m$ matrix as
\beq
\left[
\begin{array}{c}
 \Psi_{1 \, n+1} \\
 \Psi_{2 \, n+1} \\
\end{array}
\right]
=
\left[
\begin{array}{cc}
 \tz \e^{\i \a H_1^{(m)}} & \e^{\i \a H_1^{(m)}} \qa_n^{(m)} \\
 \e^{\i \a H_2^{(m)}} \ra_n^{(m)} & \frac{1}{\tz} \e^{\i \a H_2^{(m)}} \\
\end{array}
\right]
\left[
\begin{array}{c}
 \Psi_{1 \, n}   \\
 \Psi_{2 \, n}   \\
\end{array}
\right] ,
\label{scattering0}
\eeq
where $\a = \frac{\pi}{4}-\frac{1}{2} \tan^{-1} \de t$. 
To simplify the analysis, 
we introduce a gauge transformation ($l$: discrete time),
\[
\left[
\begin{array}{c}
 \Phi_{1 \, n}   \\
 \Phi_{2 \, n}   \\
\end{array}
\right]
 = 
\left[
\begin{array}{cc}
 \e^{-\i (n \a + l \be)H_1^{(m)}} &  \\
      &  \e^{-\i (n\a + l\be)H_2^{(m)}}  \\
\end{array}
\right] 
\left[
\begin{array}{c}
 \Psi_{1 \, n}   \\
 \Psi_{2 \, n}   \\
\end{array}
\right] , \hspace{5mm} \be = \tan^{-1} \de t,
\]
by which the linear problem (\ref{scattering0}) is changed into 
the standard form,
\beq
\left[
\begin{array}{c}
 \Phi_{1 \, n+1} \\
 \Phi_{2 \, n+1} \\
\end{array}
\right]
=
\left[
\begin{array}{cc}
 z I & \Q1_n^{(m)} \\
 \R1_n^{(m)} & \frac{1}{z}I \\
\end{array}
\right]
\left[
\begin{array}{c}
 \Phi_{1 \, n}   \\
 \Phi_{2 \, n}   \\
\end{array}
\right].
\label{scat_pro2}
\eeq
Here the transformed potentials are given by
\bea
\Q1_n^{(m)} = \e^{-\i(n \a  + l\be ) H_1^{(m)}} 
\qa_n^{(m)} \e^{\i (n\a  + l\be )H_2^{(m)}},
\nn \\
\R1_n^{(m)} = \e^{-\i(n\a + l\be ) H_2^{(m)}} 
\ra_n^{(m)} \e^{\i (n\a +l\be ) H_1^{(m)}}.
\nn
\eea
%
If we set 
\beq
\begin{array}{l}
\Ds 
\e^{ 2\i (n \a + l \be)} q_n^{(j)} = v_n^{(2j-2)}+ \i v_n^{(2j-1)},
\vphantom{\sum_k}
\\  \Ds
\e^{ -2\i (n \a + l \be)} r_n^{(j)} = -v_n^{(2j-2)}+ \i v_n^{(2j-1)},
\end{array}
\hspace{5mm} j= 1, 2, \ldots, m,
\label{comp}
\eeq
$Q_n^{(m)}$ and $R_n^{(m)}$ for $m\ge 2$ are expressed as
 \beq
 Q_n^{(m)}= v_n^{(0)} \eins + \sum_{k=1}^{2m-1} v_n^{(k)} e_k, 
 \hspace{5mm}
 R_n^{(m)}= -v_n^{(0)} \eins + \sum_{k=1}^{2m-1} v_n^{(k)} e_k.
\label{Q_R_const}
\eeq
Here $2^{m-1}\times 2^{m-1}$ matrices $\{e_1, \cdots, e_{2m-1} \}$ 
satisfy the following important relations:
\bea
 \{ e_i, e_j \}_+ \eq  -2 \de_{ij} \eins, 
\nn \\
 e_k^{\, \dagger} \eq - e_k, \hspace{8.5mm}
\nn
\eea
where $\{\cdot, \cdot \}_+$ denotes the anticommutator and $\eins$ is 
the $2^{m-1}\times 2^{m-1}$ unit matrix. 

It should be remarked that the transformation (\ref{comp}) 
connects the full-discrete CNLS equations (\ref{fCNLS}) with
full-discrete coupled modified KdV equations \cite{Tsuchida4,H1},
\beq
\begin{array}{l}
\Ds 
\frac{1}{\de t} (\ti{v}_n^{(j)}-v_n^{(j)})
= \Bigl(1 + \sum_{k=0}^{2m-1} v_n^{(k)\, 2} \Bigr) \Gamma_{n+1} 
(\ti{v}_{n+1}^{(j)} - v_{n-1}^{(j)}) ,
\vphantom{\sum_{\stackrel{B}{<}}}
\hspace{5mm}
j=0,1, \ldots, 2m-1,
\\ 
\Ds
\hspace{10mm}
\Gamma_{n+1} \Bigl( 1 + \sum_{k=0}^{2m-1} v_n^{(k)\, 2}\Bigr) 
= \Gamma_n \Bigl( 1 + \sum_{k=0}^{2m-1} \ti{v}_n^{(k)\, 2} \Bigr),
\end{array}
\label{fdcmkdv}
\eeq
where 
\[
\Gamma_n = \frac{\sin \be}{\de t} \ja_n . \hspace{10mm}
\]

We can investigate the direct and the inverse scattering 
problems associated with (\ref{scat_pro2}) 
in the same way as in the semi-discrete
case \cite{Tsuchida2,Tsuchida3} by assuming $m \ge 2$, the reduction 
$r_n^{(j)}= - q_n^{(j)\, \ast}$ and the boundary conditions,
\beq
q_n^{(j)}, r_n^{(j)} \to 0, \; \; 
j_n \to \frac{\de t}{\sin \be}  \; \;\; {\rm as} \; \; \;
n \to \pm \infty.
\label{bc}
\eeq
In fact, $j_n \to {\rm const.} \; (n \to \pm \infty)$ 
is sufficient to perform the ISM\@. 
We have chosen the constant as (\ref{bc}) just for convenience. 

We introduce Jost functions $\phi_n$ and $\5{\phi}_n$ as solutions 
of (\ref{scat_pro2}) which satisfy the boundary conditions:
\[
 \phi_n \sim
 \left[
 \begin{array}{c}
   I  \\
   O  \\
 \end{array}
 \right]
 z^n, 
 \hspace{5mm}
 \bar{\phi}_n \sim
 \left[
 \begin{array}{c}
   O  \\
   -I  \\
 \end{array}
 \right]
 z^{-n}
\hspace{5mm}
 {\rm as}~~~ n \rightarrow -\infty, 
\]
\[
 \phi_n \sim
 \left[
 \begin{array}{c}
   A(z) z^n \\
   B(z) z^{-n}  \\
 \end{array}
 \right],
 \hspace{5mm}
 \bar{\phi}_n \sim
 \left[
 \begin{array}{c}
  \5{B}(z) z^n \\
  -\5{A}(z) z^{-n}  \\
 \end{array}
 \right]
\hspace{5mm}
 {\rm as}~~~ n \rightarrow +\infty.
\]
Here $\{$$A(z)$, $\5{A}(z)$, $B(z)$, $\5{B}(z)$$\}$ 
are $n$-independent $2^{m-1} \times 2^{m-1}$ matrices which 
are called scattering data. 
Suppose that $1/\det A(z)$ has $2N$ isolated simple
 poles $\{z_1, z_2, \cdots, z_{2N} \}$ in $|z|>1$ and is regular on the unit 
circle $C$. It will be explained why we choose the number of poles 
to be even. Let us define $F(n)$ in terms of the scattering data by 
\bea
 F(n) \eq \frac{1}{2\pi \i} \oint_{C} B(z)A(z)^{-1} z^{-n-1} \d z
+ \sum_{j=1}^{2N} C_j z_j^{-n-1},
\nn 
\eea
where $C_j$ is the residue matrix of $B(z) A(z)^{-1}$ at $z = z_j$. 
Similarly, we suppose that $1/\det \5{A}(z)$ has $2\5{N}$ isolated simple
 poles $\{\5{z}_1, \5{z}_2, \cdots, \5{z}_{2\5{N}}
 \}$ in $|z|<1$ and is regular on the unit circle $C$. We define 
$\5{F}(n)$ by
\bea
 \5{F}(n) \eq 
\frac{1}{2\pi \i} \oint_{C} 
\bar{B}(z)\bar{A}(z)^{-1} z^{n-1} \d z 
 - \sum_{k=1}^{2 \5{N}} \5{C}_k \5{z}_k^{n-1},
\nn
\eea
where $\5{C}_k$ is the residue matrix of $\5{B}(z) \5{A}(z)^{-1}$
 at $z = \5{z}_k$. It can be shown after some computation that 
$F(n)$ and $\5{F}(n)$ are connected with the physical variables 
$Q_n^{(m)}$ and $R_n^{(m)}$ through the following
Gel'fand-Levitan-Marchenko equations:
\bea
\kappa_{1} (n,q) 
\eq  \5{F} (n+q) 
- 
       \sum_{n'=n+1}^\infty 
        \sum_{n''=n+1}^\infty 
         \kappa_{1 } (n, n'') F (n''+n') \5{F} (n'+q),
\label{GLL1} 
\\
%
\5{\kappa}_{2 } (n,q) 
\eq  - F (n+q) 
      - \sum_{n'=n+1}^\infty 
        \sum_{n''=n+1}^\infty 
         \5{\kappa}_{2 } (n, n'') \5{F}(n''+n') F(n'+q),
\label{GLL2}
\eea
for $q>n$. Here
%
 \[
 -\kappa_1(n,n+1) = Q_n^{(m)}, \hspace{5mm}
 -\bar{\kappa}_2(n,n+1) = R_n^{(m)}.
 \]

It should be noted that the scattering problem 
(\ref{scat_pro2}) gives the symmetry properties 
of the scattering data. 
Iterating (\ref{scat_pro2}), we can prove that 
$A(z)$, $\5{A}(z)$ are polynomials in $z$ of even degree and 
$B(z)$, $\5{B}(z)$ are polynomials in $z$ of odd degree. This 
fact gives the symmetry properties of $F$ and $\5{F}$:
\[
F(n) = 
\left\{
\begin{array}{c}
 2 F_R (n), \\
  O, \\
\end{array}
\right.
\begin{array}{l}
 n: {\rm odd},  \\
 n: {\rm even}, \\
\end{array}
\]
%
\[
\5{F}(n) = 
\left\{
\begin{array}{c}
 2 \5{F}_R (n), \\
  O, \\
\end{array}
\right.
\hspace{2mm}
\begin{array}{l}
 n: {\rm odd}, \\
 n: {\rm even}. \\
\end{array}
\]
Further, due to the internal symmetries of $Q_n^{(m)}$ and $R_n^{(m)}$ 
given by (\ref{Q_R_const}), $F_R(n)$ and $\5{F}_R(n)$ for odd $n$ are 
expressed as
\bea
F_R(n) = f^{(0)}(n)\eins + \sum_{k=1}^{2m-1}f^{(k)}(n) e_k,
\hspace{5mm}
%
\5{F}_R(n) = f^{(0)}(n)\eins - \sum_{k=1}^{2m-1}f^{(k)}(n) e_k.
\nn
\eea
Here the functions $f^{(0)}(n)$ and $f^{(k)}(n)$ are real. 
From the asymptotic form of the Lax matrix $V_n$ as $n \to \pm
\infty$, the time dependences of the scattering data are given by
%
%
\bseq
\beq
A(z,l) = A(z,0),
\eeq
\beq
B(z,l) A(z,l)^{-1} 
= B(z,0)A(z,0)^{-1} \Biggl( \frac{1-\de t \, z^2}{1-\de t/z^2}
\Biggr)^l , 
\eeq
\beq
C_j(l)= C_j(0)\Biggl( \frac{1-\de t \, z_j^2}{1-\de t/z_j^2} \Biggr)^l.
\eeq
\label{time_dep}
\eseq
Thus $f^{(0)}(n)$ and $f^{(k)}(n)$ satisfy the following dispersion 
relation:
\[
 \frac{1}{\de t} \{  \ti{f}^{(j)}(2n+1) - f^{(j)}(2n+1) \}
= \ti{f}^{(j)}(2n+3) - f^{(j)}(2n-1), \hspace{5mm}
j=0, 1, \ldots, 2m-1.
\]
%
%

We can solve the initial-value problem of the full-discrete 
CNLS equations (\ref{fdcH}) in the following steps. 
\begin{enumerate}
\item[(a)]
For given potentials at $l=0$, $Q^{(m)}_n(0)$ and $R^{(m)}_n (0)$, 
we solve the direct problem of scattering, (\ref{scat_pro2}), and obtain 
the scattering data $ \{ B(z)A(z)^{-1}, z_j, C_j\}$.
\item[(b)]
The time dependences of the scattering data are given by (\ref{time_dep}).
%
\item[(c)]
We substitute the time-dependent scattering data into 
the Gel'fand-Levitan-Marchenko equations (\ref{GLL1}) and
(\ref{GLL2}). 
Solving the equations, we reconstruct the time-dependent potentials,
 \[
 Q^{(m)}_n (l) = -\kappa_1(n,n+1;l), \hspace{5mm} 
 R^{(m)}_n (l) = -\bar{\kappa}_2(n,n+1;l).
 \]
This step corresponds to solving the inverse problem of scattering.
\end{enumerate}

%
In order to obtain soliton solutions, we assume 
that each soliton seen in $\sum_j |q_n^{(j)}(l)|^2$ has a
time-independent shape. By calculating an asymptotic behavior of the
tails of solitons at $n \, \to +\infty$, we obtain the corresponding
conditions, 
\[
C_{2j-1} \bar{C}_{2j} = \bar{C}_{2j} C_{2j-1}
 = C_{2j} \bar{C}_{2j-1} = \bar{C}_{2j-1} C_{2j} = O,
\hspace{10mm} j=1, 2, \ldots, N,
\]
on the residue matrices.
%
For instance, the one-soliton solution 
of (\ref{fdcH}) is computed as 
\bea
q_n^{(i)}(l) \eq
\sech \bigl\{ 2 n W + 2 l a + \phi_0 \bigr\} 
   \frac{\sinh 2W}{\sqrt{\displaystyle \sum_{j=1}^{m}
(|\alpha_j|^2 +|\beta_j|^2)}}
\bigl[ \alpha_i \e^{2\i \{n (\theta - \a) 
   + l (b-\beta) \}}
\nn \\  
&&
\mbox{} + \beta_i^\ast \e^{-2\i \{n (\theta + \a) 
   + l(b+\beta) \}} 
  \bigr] , \; \; \;
\; \; \;  i=1, 2, \ldots, m ,
\nn
\eea
where $W > 0$, $\Ds \sum_{i=1}^{m} \alpha_i \beta_i =0$ and
\bea
\hspace{-7.5mm}
\e^{\phi_0} \eq
\frac{\sinh 2W}{\sqrt{\displaystyle \sum_{j=1}^m
(|\a_j|^2 + |\be_j|^2)}} ,
\nn 
\\
\nn \\
\hspace{-7.5mm}
\e^{4a} \eq \frac{(1-\de t \, \e^{-2W - 2\i\theta})(1-\de t \,
      \e^{-2W+2\i \theta})}
{(1-\de t \, \e^{2W+ 2\i \theta})(1-\de t \, \e^{2W-2\i \theta})},
%
\nn 
\\
\nn \\
\hspace{-7.5mm}
\e^{4\i b} \eq  \frac{(1-\de t \, \e^{2W - 2\i\theta})(1-\de t \,
      \e^{-2W-2\i \theta})}
{(1-\de t \, \e^{2W+ 2\i \theta})(1-\de t \, \e^{-2W+2\i \theta})}.
\nn
\eea
In the continuum limit of time ($\de t \to 0$), the one-soliton 
solution reduces to that of the semi-discrete CNLS 
equations \cite{Tsuchida3} due to the relations,
\bea
a \eq \de t \sinh 2W \cos 2 \theta + O(\de t^2),
\nn \\
b \eq \de t \cosh 2W \sin 2\theta + O(\de t^2).
\nn
\eea
It is noteworthy that we can obtain the most general soliton solutions 
of the model since we have employed the ISM for the analysis. 

Conserved quantities of the full-discrete CNLS equations (\ref{fCNLS}) 
under the decaying boundary conditions 
can be computed recursively by expanding the time-independent quantity, 
$\log \det A(z)$, with respect to $1/z$. 
Here we list first three conserved densities,
%
\[
\log \Bigl( 1-\sum_j q^{(j)}_n r^{(j)}_n \Bigr),
\]
\[
\cos 2\a \sum_j (q^{(j)}_{n+1} r^{(j)}_n + q^{(j)}_{n} r^{(j)}_{n+1})
+ \i \sin 2\a \sum_j
 (q^{(j)}_{n+1} r^{(j)}_n - q^{(j)}_{n} r^{(j)}_{n+1}), 
\]
\bea
&&
\hspace{-5mm}
\Bigl( 1 - \sum_{j} q^{(j)}_{n+1} r^{(j)}_{n+1} \Bigr)
\Bigl\{ \cos 4\a 
\sum_{j} ( q^{(j)}_{n+2} r^{(j)}_n + q^{(j)}_{n} r^{(j)}_{n+2} )
+\i \sin 4\a
\sum_{j} ( q^{(j)}_{n+2} r^{(j)}_n - q^{(j)}_{n} r^{(j)}_{n+2} )
\Bigr\}
\nn \\
&&
\hspace{-5mm}
\mbox{} -\frac{1}{2} \Bigl\{ \cos 2\a 
\sum_{j} (q^{(j)}_{n+1} r^{(j)}_n + q^{(j)}_n r^{(j)}_{n+1}) 
+\i \sin 2\a
\sum_{j} (q^{(j)}_{n+1} r^{(j)}_n - q^{(j)}_n r^{(j)}_{n+1}) 
\Bigr\}^2 
\nn \\
&&
\hspace{-5mm}
\mbox{} + \sum_{j} q^{(j)}_{n+1} r^{(j)}_{n+1}
        \sum_{j} q^{(j)}_{n} r^{(j)}_{n} .
\nn
\eea
%

\section{Concluding remarks}
\label{}

We have investigated an integrable full-discretization of 
the coupled nonlinear Schr\"{o}dinger (CNLS) equations 
from the point of view of the 
inverse scattering method (ISM) for the first time in this article. 
We have found a Lax pair for full-discrete CNLS equations and
performed the ISM under appropriate conditions with the help of the 
transformation (\ref{comp}). As a result, we can solve the 
initial-value problem, and obtain soliton solutions and conserved
quantities. 

There are plural schemes for integrable full-discretization of
one-component nonlinear Schr\"{o}dinger (NLS) equation. We have some 
freedom to determine the linearized dispersion relation of
full-discrete NLS equation (see \cite{AL2}). However, as far as we
have considered, there is no such freedom in the multi-component case, 
which leads us to (\ref{fCNLS}) as full-discrete CNLS equations. 
It is to be noted that (\ref{fCNLS}) can be cast into a few types of 
full-discrete CNLS equations by coordinate transformations and
redefinitions of the auxiliary variable $\ja_n$ (see \cite{Sasa} for 
the work of Hirota, Ohta and Tsujimoto on full-discrete NLS equation).

Finally, we comment that 
discrete CNLS equations 
have been studied by using the direct method \cite{H1,Ohta2,Ohta3}. 
In \cite{H1}, Hirota proposed the full-discrete coupled modified 
KdV equations (\ref{fdcmkdv}) and gave a $(1+1)$-soliton solution 
in the two-component case. Ohta obtained an $N$-soliton solution 
for the semi-discrete CNLS equations in terms of 
Pfaffian \cite{Ohta2}. However, the $N$-soliton solution is not 
a general one since the initial soliton polarizations in the solution 
are either parallel or orthogonal. More recently, 
by using 
another Pfaffian representation, 
Ohta proposed more general soliton solutions 
without a restriction on soliton polarizations in the full-discrete
case \cite{Ohta3}. He also obtained a Lax pair of different form 
in \cite{Ohta3} by means of an alternative approach.

\end{document}